\documentclass[12pt,a4paper]{article} 

\pdfoutput=1
\usepackage[numbers,sort&compress]{natbib}
\usepackage[affil-it]{authblk}

\usepackage{amsmath}
\usepackage{amsthm}
\usepackage[english]{babel}
\usepackage{graphicx}
\usepackage{amsfonts} 
\usepackage{amssymb} 
\usepackage{hyperref} 
\usepackage{float}
\usepackage{times}
\usepackage{tensor}
\usepackage{mathtools}
\usepackage{color}

\numberwithin{equation}{section}

\newcommand{\deriv}[1]{\frac{d}{d #1}}
\newcommand{\DERIV}[2]{\frac{d #1}{d #2}}
\newcommand{\pd}[1]{\frac{\partial}{\partial #1}}
\newcommand{\PD}[2]{\frac{\partial #1}{\partial #2}}

\newcommand{\half}[0]{\frac{1}{2}}

\newcommand{\oppr}[3]{\ensuremath{\left \langle \left. #1\right.  \right| #2 \left| \left. #3 \right. \right \rangle}}

\title{Classical and quantum cosmology with York time} 
\author{Philipp Roser and Antony Valentini}
\affil{Department of Physics and Astronomy,\\ Clemson University, Kinard Laboratory,\\ Clemson, SC 29631-0978, USA}
\date{\vspace{-0.5cm}}

\begin{document}
 \maketitle
 \begin{abstract}
 We consider a solution to the problem of time in quantum gravity by deparameterisation of the ADM action in terms of York time, a parameter proportional to the extrinsic curvature of a spatial hypersurface. We study a minisuperspace model together with a homogeneous scalar field, for which we can solve the Hamiltonian constraint exactly and arrive at an explicit expression for the physical (non-vanishing) Hamiltonian. The scale factor and associated momentum cease to be dynamical variables, leaving the scalar field as the only physical degree of freedom. We investigate the resulting classical theory, showing how the dynamics of the scale factor can be recovered via an appropriate interpretation of the Hamiltonian as a volume. We then quantise the system in the Schr\"odinger picture. In the quantum theory we recover the dynamics of the scale factor by interpreting the spectrum and expectation value of the Hamiltonian as being associated with volume rather than energy. If trajectories in the sense of de~Broglie-Bohm are introduced in the quantum theory, these are found to match those of the classical theory. We suggest that these trajectories may provide the basis for a perturbation theory in which both background and perturbations are quantised.
\end{abstract}

\section{Introduction}

Many of the difficulties encountered when attempting to construct a quantum theory of gravity have their origin in what has been broadly labelled as `the problem(s) of time', most prominently among them the `frozen' dynamics implied by the Wheeler-deWitt equation. A large literature has been dedicated to discussing these issues and attempts to resolve them (see \citep{Kuchar2011,Rovelli2007,Thiemann2008,Anderson2012} and references therein).

One possibility is the identification of a `physical time', that is, a physically preferred temporal parameter, which may be extrinsic (derived from the gravitational/ geometric degrees of freedom including the extrinsic curvature), intrinsic (taken from among the gravitational/geometric 3-variables alone), or derived from non-gravitational degrees of freedom such as matter fields. For an example of `matter time', see \citep{BrownKuchar1995}, \citep{HusainPawlowski2012}. In this paper we will consider an example of extrinsic time.

The question arises how a time parameter should be identified from among the gravitational degrees of freedom. Furthermore, the choice made must be justified if it is to be understood as physically preferred rather than just computationally convenient. Choosing such a time parameter is equivalent to choosing a 3+1-slicing of the spacetime based on a geometric property of the slices. It is crucial to note that while in the classical theory one may disregard such choices of time parameter as a mere gauge-fixing with identical phenomenology, in general this need not be true following quantisation. That is, different choices of slicing followed by quantisation can in principle lead to different phenomenologies. Thus quantum-gravitational effects might, at least in principle, help identify true physical time if indeed there is one.

One suggestion \citep{ChoquetBruhatYork1980} has been to identify the trace $K$ of the extrinsic curvature $K^{ij}$ with time, leading to spatial slices of constant mean curvature (CMC), parametrised by `York time' $T=\frac{1}{12\pi G}K$. As we shall discuss in section \ref{YorkTimeBackground}, for a homogeneous expanding universe York time is proportional to (the negative of) the Hubble parameter. Thus in a sense $T$ may be regarded as a measure of spatial expansion.

The importance of CMC slices has been known at least since York's solution of the initial-value problem of general relativity \citep{York1972,York1973}. The initial-value problem is not straightforward due to the requirement that the constraints arising from the four-diffeomorphism invariance must be satisfied, yet sufficient data must be provided that the future evolution is uniquely determined (up to local gauge transformations). York showed that the specification of the conformal part of the three-metric and associated momentum on a CMC slice fulfils this requirement, where the CMC property allows a decoupling of the Hamiltonian and momentum constraints. 

One might consider York's method not only a convenient mathematical tool for finding consistent initial conditions, but also an indication that only those degrees of freedom are, in fact, physical. Recently this line of enquiry has inspired a theory of gravity based on a conformal 3+1-formalism that is phenomenologically equivalent to general relativity (up to the exclusion of certain pathological solutions of GR) known as shape dynamics \citep{GomesGrybKoslowski2011,Mercati2014}, based on earlier work on conformal theories of gravity by Barbour and collaborators (for example, \citep{AndersonBarbourFosterOMurchadha2003,AndersonBarbourFosterKelleherOMurchadha2005}). Here the problem of time is addressed by moving to entirely dimensionless variables. The natural temporal parameter is the (also dimensionless) ratio of the York parameter at one instant and at an arbitrarily chosen reference time \citep{BarbourKoslowskiMercati2013ProbOfTime}. The value of this parameter is inaccessible, but a dimensional 'time' as measured by physical clocks can be recovered entirely within the gravitational dynamics \citep{BarbourKoslowskiMercati2013GraviArrow}. The approach taken in the present paper is more standard in that we start from general relativity and a 3+1 splitting rather than from an evolving 3-dimensional conformal geometry. Ultimately there may be good reasons to prefer the shape dynamics starting point. However, here we start with cosmology in the ADM description.

Another compelling reason for investigating gravitational dynamics with $T$ as a physical time parameter has been given by one of the authors \citep{Valentini1996,Valentini1997,Valentini2008inCraigSmith}. Here it was argued that in the de~Broglie-Bohm formulation of quantum theory \citep{deBroglie1928,Bohm1952a,Bohm1952b,BacciagaluppiValentini2009,Holland1993} and other hidden-variable approaches a preferred slicing of spacetime is required in order to make sense of non-locality. Non-local mechanisms in an arbitrary slicing would lead to inconsistencies. It was suggested that York time has certain properties that make it suitable as a physical time parameter. Thus there are reasons on both the classical and quantum level to consider York time as an absolute time parameter.

We wish to point out that there are, however, other options. An interesting suggestion has been made by Smolin and Soo \citep{SmolinSoo1995}, who argue that the Chern-Simons invariant also has properties that make it viable as a physical time coordinate. In fact, the authors show that in an appropriate limit the Chern-Simons invariant becomes imaginary and reduces to the extrinsic curvature. 

In this paper we focus on carrying out a formal quantisation, without committing ourselves to a particular understanding of what is meant by 'time'. Broadly speaking there are two main approaches, on which we shall briefly comment.

The first, `pre-relativistic' approach regards time as an external parameter. On this view, time is a primitive and need not be thought of as defined operationally (for example by a clock). One may regard the time parameter as defined in such a way as to make the equations of motion as simple as possible (see for example \citet[pp.\ 23-27]{MisnerThorneWheeler1973}). While this approach fell out of favour with the development of relativity theory, in which there is no preferred frame, in our view it is not necessarily problematic; and it could be advantageous in the context of quantum gravity, in particular in a de Broglie-Bohm formulation \citep{Valentini1996,Valentini1997,Valentini2008inCraigSmith}.

The second, `operational' view regards time as necessarily defined in terms of clocks or other material systems. If one adopts the second viewpoint, we suggest that a natural approach would be to operationally identify time with the expansion of space. While a proper analysis lies outside the scope of this paper, in effect we are suggesting that measurements of the Hubble parameter could be reinterpreted as measurements of York time (at least for the case of a spatially homogeneous universe). Alternatively, an operational approach might best be implemented in terms of the dimensionless analogue of York time employed by \citet{BarbourKoslowskiMercati2013ProbOfTime}.

The method we pursue here to obtain the dynamics for a minisuperspace model with York time is via a reduced Hamiltonian formalism. Realising that the Einstein-Hilbert action, when decomposed into 3+1-form, is in parameterised form \citep{ADM1962}, we identify one local degree of freedom $q_0$ to serve as time and solve the Hamiltonian constraint $\mathcal{H}$ for the momentum $P_{q_0}$ conjugate to $q_0$. The physical (non-vanishing) Hamiltonian of our deparameterised theory is then given by $H_{q_0}\equiv -P_{q_0}$. The physical Hamiltonian $H_{q_0}$ is now given as a function of the remaining degrees of freedom and possibly $q_0$ itself, that is, it may carry an explicit time dependence. The physical meaning of the numerical value of $H_{q_0}$ results from the choice of $P_{q_0}$ when deparameterising. In a parameterised particle model the momentum conjugate to Newtonian time $t$ is $-E$, where $E$ is the energy of the system (see \citet{Lanczos1949}). Deparameterisation to Newtonian time then yields a Hamiltonian whose numerical value is the energy. If the solution, that is, the expression for $H_t$ in terms of the other degrees of freedom, happens to be time-independent, then energy will be conserved. In the case of York time $T$, the reduced Hamiltonian density turns out to be the local volume element $\sqrt{g}$, where $g$ is the determinant of the local 3-metric $g_{ab}$.\footnote{For a general foliation of space-time into 3-space slices, there is of course nothing that guarantees the constancy of $T$ across each slice. This must therefore be imposed as a condition on the slicing, enforcing constant mean curvature across each slice (see section \ref{YorkTimeBackground}). The Hamiltonian will then be an integral of $-P_T$ over the slice, although in a minisuperspace model this simply gives a constant and no spatial integral appears.} The general consistency of the reduced Hamiltonian procedure for the ADM action has been known since \citep{ADM1960}. 

This approach has been considered by Choquet-Bruhat and York \citep{ChoquetBruhatYork1980}, but solving the Hamiltonian constraint for $P_T$ is difficult and an analytic solution for the general case is unknown. Following a short review of York time (section \ref{YorkTimeBackground}), in this paper we will perform the procedure explicitly in the context of a minisuperspace model with a homogeneous scalar field $\phi$. To the authors' knowledge this dynamics has never been developed fully for a simple model such as the one presented here, although some formal results exist (for example \citep{Jantzen1980}) and York time has been repeatedly considered throughout the literature. As we will show, the Hamiltonian constraint takes the form of a depressed cubic that may be readily solved. In the case of a spatially flat universe the resulting reduced Hamiltonian for $\phi$, now the homogeneous field value across all space, turns out to be linear in momentum $p_\phi$ and furthermore explicitly time-dependent (section \ref{ReducedHamiltonian}). The dynamics for both zero and non-zero potential $V(\phi)$ is governed by first-order equations (section \ref{Classical}). The dynamics is equivalent to the standard Friedmann-Lema\^itre and Klein-Gordon equations, as is shown in the appendix. This reduction of order as compared to the cosmological-time dynamics ---at the cost of the introduction of an explicit time dependence--- is a general feature of York time.

We then quantise the theory in the Schr\"odinger picture and analyse the resulting dynamics (section \ref{Quantum}). It turns out to be useful to perform the analysis of the quantum theory in the de~Broglie-Bohm formulation, although we emphasise that our quantisation procedure is independent of doing so.\footnote{The de~Broglie-Bohm formulation has been used in quantum cosmology by a number of authors, such as by Pinto-Neto and collaborators in connection with bouncing models \citep{PeterPinhoPintoneto2007,Pinto-NetoSantosStruyve2014}, by Pinto-Neto, Santos and Struyve \citep{Pinto-NetoSantosStruyve2012} and Valentini \citep{Valentini2010InflCosm} in relation to inflation, and most recently by He, Gao and Cai in work on the mathematical foundations of big-bang cosmology \citep{HeGaoCai2014}.} The interesting result obtained here is that, when calculating quantum trajectories as postulated by the de~Broglie-Bohm approach, we find that they exactly match those of the classical theory. In the theory with non-zero potential a factor ordering ambiguity arises, although a symmetric ordering is shown to have unique properties. If one considers the conventional quantum theory for this system (without trajectories), the corresponding result is that the probability density evolves like a classical ensemble. An implication to be explored in future work is that, if one considers our minisuperspace model as a `background' on which to introduce perturbations, one can treat both background and perturbations quantum-mechanically, that is, one does not have to rely on a `classical' background on which only perturbations are quantised.

The explicit time dependence of the Hamiltonian implies that its eigensolutions vary in time. In analogy with energy in more conventional systems, we identify the eigenvalues of the Hamiltonian as `volume eigenvalues' and the corresponding eigenstates as `volume eigenstates'. We show that for the free theory the fractional rate of change of the volume eigenvalues in time is universal across all eigensolutions. Volume expectation values therefore all share the same fractional rate of change, which matches the classical one for the free theory, where one may thus recover the Friedmann-Lema\^{i}tre dynamics entirely on the quantum level. In the case of non-zero potential a difficulty arises in that the eigenfunctions are not normalisable. We compute an expression for the fractional rate of change of volume, although in practice some form of regularisation method will have to be used to interpret this. In the case of non-zero $V(\phi)$ the eigenfunctions themselves are furthermore time-dependent, which leads to a state-dependent term in the computed expression.

We conclude with a discussion and suggestions for future work.

A note on conventions: Throughout this article Latin indices are understood to take values $1,2,3$, ranging over spatial dimensions only, while Greek indices also include the temporal direction, taking values $0,1,2,3$. We use a 4-metric signature of $-+++$. The left-superscript $^{(4)}$ signifies 4-quantities in space-time. We use units such that $c=\hbar=1$, but refrain from normalising the gravitational constant $G$ since we consider it useful to see how the dynamics depends on the physical value of $G$.

\section{Review of York time}\label{YorkTimeBackground}

\subsection{Extrinsic curvature and York time}
The extrinsic curvature $K_{ij}$ of a spatial hypersurface describes the manner in which the slice is embedded in the full space-time manifold. Thus an isolated slice cannot be said to have an extrinsic curvature. Formally $K_{ij}$ may be defined in terms of the Lie derivative of the spatial 3-metric $g_{ij}$ on the slice in the direction $\hat{n}$ that is normal to the hypersurface:
\begin{equation} K_{ij}:=-\half \mathcal{L}_{\hat{n}}g_{ij}=-\half\mathcal{L}_{\hat{n}} \prescript{(4)}{}g_{ij}
			 =-\half \left(n^\mu\partial_\mu ^{\;(4)}g_{ij}+ \prescript{(4)}{}g_{i\mu}\partial_jn^\mu+\prescript{(4)}{}g_{\mu j}\partial_in^\mu\right). 
\end{equation}
$K_{ij}$ contains information about the parallel-transport properties of the normal vector $n_i$ (as well as of vectors tangent to the hypersurface) from one slice to another. It may be re-expressed in terms of the covariant derivative as $K_{ij}=-\nabla_jn_k$.

In order to describe the embedding of a slice with respect to others one splits the 4-metric $^{(4)}g_{\mu\nu}$ into its time-time, space-time and space-space parts using the lapse and shift functions $N$ and $N^i$ together with the 3-metric $g_{ij}$ defined on the slice \citep{ADM1962}. A space-time interval now takes the form
\begin{equation} ds^2 = g_{\mu\nu}dx^\mu dx^\nu = -(N^2-N_iN^i)dt^2+N_idx^idt+g_{ij}dx^idx^j. \end{equation}
The lapse $N$ can be understood to give the rate of change of proper time with respect to coordinate time in the direction normal to the slice, $d\tau=Ndt$, while the shift $N^i$ determines the deviation of lines of constant spatial coordinate from the normal 4-vector $n^\mu$ to the spatial slice. A proper 4-volume element is given by $\sqrt{\prescript{(4)}{}g}=N\sqrt{g}$, where $g=\det(g_{ij})$. For a detailed discussion of the 3+1-slicing in this manner, as well as for the formal properties of the extrinsic curvature $K_{ij}$, see \citep{MisnerThorneWheeler1973}.

In terms of the lapse $N$ and shift $N^i$, $K_{ij}$ takes the form
\begin{equation} K_{ij}=\frac{1}{2N}\left(\nabla_iN_j+\nabla_jN_i-\PD{g_{ij}}{t}\right),\end{equation}
where $t$ is an arbitrary temporal parameter labelling the slices and $\nabla_i$ is the spatial covariant derivative on the slice. For the remainder of this paper we will set the shift $N_i$ to zero everywhere. This is always possible provided lines of constant spatial coordinates do not reach a singularity. Setting $N^i=0$ can be understood as a partial gauge fixing of the 4-diffeomorphism invariance of general relativity.

The idea of using the scalar extrinsic curvature $K\equiv g^{ij}K_{ij}$ as a time parameter originated in York's work on the initial-value problem of general relativity \citep{York1972,York1973}, that is, the problem of finding the most general complete set of initial data on a space-like hypersurface such that the Hamiltonian and momentum constraints are satisfied. The constraints are
\begin{align}
0 &= \mathcal{H}^a \equiv-2\nabla_bp^{ab} \label{momentumconstraint} \\ 
0 &= \mathcal{H}   \equiv\frac{16\pi G}{\sqrt{g}}\left(p_{ab}p^{ab}-\frac{1}{2}p^2\right)-\frac{\sqrt{g}}{16\pi G}R, \label{Hamiltonianconstraint}
\end{align}
where $p^{ab}$ is the momentum conjugate to $g_{ab}$ and $p=g_{ab}p^{ab}$ denotes its trace.

York's insight was to begin by constructing a pair of symmetric $g_{ab}$ and $p^{ab}$ which satisfy the momentum constraint (\ref{momentumconstraint}), which is easily done. On a maximal hypersurface ($K=0$) the momentum constraint is preserved under conformal transformations $g_{ab}\rightarrow \bar{g}_{ab}=\varphi^4g_{ab}$, $p^{ab}\rightarrow \bar{p}^{ab}=\varphi^{-4}p^{ab}$ (the power of 4 is just for convenience). Performing such a transformation implies that the Hamiltonian constraint (\ref{Hamiltonianconstraint}) is satisfied if and only if the conformal factor $\varphi$ satisfies 
\begin{equation} \frac{1}{16\pi G}(-8\nabla^2\varphi + R\varphi) = 16\pi G g^{-1} p_{ab}p^{ab}\varphi^{-7}, \label{Lichnerowicz} \end{equation}
known as the Lichnerowicz equation. $R$ is the spatial scalar curvature. Then the pair $\bar{g}_{ab}$, $\bar{p}^{ab}$ satisfies both constraints as was required. The procedure is extended to non-maximal but constant-mean-curvature hypersurfaces by replacing $p^{ab}$ by its trace-free part $\sigma^{ab}:=p^{ab}-\frac{p}{3}g^{ab}$, which is transverse on CMC surfaces. The Lichnerowicz equation then takes the form
\begin{equation} \frac{1}{16\pi G}(-8\nabla^2\varphi+R\varphi) = 16\pi G\left( g^{-1}\sigma_{ab}\sigma^{ab}\varphi^{-7}-\frac{3}{8}T^2\varphi^5\right). \label{ExtendedLichnerowicz} \end{equation}
The above equation holds for the vacuum case. The inclusion of general matter terms is a subtle issue in that the power $r$ in the rescaling factor $\varphi^r$ appropriate for the energy density depends on whether or not this source carries a fundamental (fixed) length scale (see \citet{ChoquetBruhatYork1980}, section 3.3.3).

The variable $T:=\frac{2p}{3\sqrt{g}}$, being proportional to $K$, is constant across each slice. It follows from the equations of motion that $T$ increases in   
positive timelike directions, and so it represents a suitable choice of time parameter. The variable conjugate to $T$ is $\sqrt{g}$, which in the reduced Hamiltonian formalism below will therefore be identified with $-H$, $H$ being the physical Hamiltonian associated with time $T$. 

A reader unfamiliar with York time may find it helpful to obtain a more intuitive understanding of the meaning of $T$. One way to characterise the extrinsic curvature is as a complete description of the deformation of a local volume element as one passes from one slice to another. The trace $K$ in particular denotes the fractional rate of change in volume, that is, $d\sqrt{g}=-K\sqrt{g}d\tau$, where $d\tau$ is an infinitesimal interval of proper time. In a cosmological scenario we have $\sqrt{g}\sim a^3$, $a$ being the global scale factor, giving $T=\frac{1}{12\pi G}K=-\frac{1}{4\pi G}\frac{\dot{a}}{a}$, so $T$ is just proportional to the negative of the Hubble parameter. The numerical constant is chosen such that the momentum conjugate to $T$ is just $P_T=-\sqrt{g}=-a^3$, where `conjugate' is defined with respect to the Einstein-Hilbert action in 3+1 ADM form.

Thus for a closed universe $T$ takes values $-\infty$ at the big bang and $+\infty$ at the big crunch. In the flat and open case $T$ begins at $-\infty$ and approaches $0$ from below as standard cosmological time $t$ approaches infinity. 

\subsection{The Friedmann-Lema\^itre equations in York time}
Consider now such a homogeneous isotropic cosmological model. We will develop this properly below, but in this section we simply wish to gain a preliminary understanding of the type of dynamics we might expect when using York time. For simplicity, in this paper we will not discuss the impact of a non-zero cosmological constant, although this may be obtained by simply choosing $V(\phi)$ to contain an additive non-zero constant. The standard form of the Friedmann-Lema\^itre equations (FLE) is
\begin{align}
 \frac{\dot{a}^2}{a^2}+\frac{k}{a^2}&=\frac{8\pi G}{3}\rho \label{CosmoFriedmann1}\\
 \frac{\ddot{a}}{a}&=-\frac{4\pi G}{3}(\rho+3\mathcal{P}). \label{CosmoFriedmann2}
\end{align}
The first of these corresponds to the time-time component of the Einstein equations and the second to a combination of the space-space and time-time components. Here a dot denotes differentiation with respect to cosmological time $t$, with $N=1$. The parameter $k$ takes values $0,+1,-1$ for a flat, closed and open universe respectively. $\rho$ denotes the energy density and for a scalar field takes the form $\rho=\half\dot{\phi}^2+V(\phi)$. The pressure $\mathcal{P}$ is given by $\mathcal{P}=\half\dot{\phi}^2-V(\phi)$.

Using $T=-\frac{1}{4\pi G}\frac{\dot{a}}{a}$, written in terms of York time equation (\ref{CosmoFriedmann1}) becomes
\begin{equation} 16\pi^2G^2T^2+\frac{k}{a^2} = \frac{8\pi G}{3}\rho, \label{GeneralFriedmann1}\end{equation}
which is purely algebraic if $\rho$ does not contain any time-derivatives, as is the case for a dust-dominated universe, or if we have a field for which appropriate ``slow-roll'' conditions are satisfied (so that we can ignore the kinetic term). This possible reduction of order in return for explicit time dependence will be a key feature we encounter in the reduced Hamiltonian formalism below. Note though that if $\rho$ does contain time-derivatives as it is the case for a general scalar field, then we must convert the derivative with respect to $t$ into one with respect to $T$ in order to avoid reference to $t$. In general this requires us to determine the York lapse $N_T$, defined by $dt=N_TdT$ (since $N_t=1$). 

Let a prime ($^\prime$) denote differentiation with respect to $T$. We have $N_T^{-1}a^\prime=\dot{a}$, which together with $T=-\frac{1}{4\pi G}\frac{\dot{a}}{a}$ implies $N_T^{-1}=-4\pi G T\frac{a}{a^\prime}$, allowing us to replace $\dot{\phi}$ in $\rho$ by $N_T^{-1}\phi^\prime$. Since $a^\prime$ appears in the lapse, the resultant form of the first FLE is not algebraic. In the simplest case, namely a free scalar field ($V=0$) in a flat ($k=0$) universe, the FLE takes the form $\frac{\dot{a}^2}{a^2}=\frac{8\pi G}{6}\dot{\phi}^2$, which gives 
\begin{equation}\frac{a^{\prime2}}{a^2}=\frac{8\pi G}{6}\phi^{\prime2} \label{freeFriedmann1}\end{equation} 
after multiplying both sides by the $N_T^{-2}$. Here explicit knowledge of $N_T$ would not have been necessary.

The simplest way to transform the second FLE into a form without reference to $t$ is to use 
\begin{equation}N_T^{-1}=\DERIV{T}{t}=-\deriv{T}\frac{1}{4\pi G}\frac{\dot{a}}{a}=-\frac{1}{4\pi G}\frac{\ddot{a}}{a}+4\pi GT^2 \end{equation}
and combine this with the previous expression for $N_T$ to give
\begin{equation}\frac{\ddot{a}}{a}=16\pi^2G^2\left(T^2+\frac{aT}{a^\prime}\right). \label{adoubledot}\end{equation}
 For the scalar field eq.\ (\ref{CosmoFriedmann2}), after some appropriate rearrangement, now takes the form
\begin{equation} \frac{a^\prime}{a}=-\frac{T}{T^2+\frac{1}{12\pi G}(\rho+3\mathcal{P})}, \label{GeneralFriedmann2}\end{equation}
which is a first order equation even after replacement of $\dot{\phi}$ by $N_T^{-1}\phi^\prime$. Here the reduction of order in return for explicit time dependence is general, provided $\rho$ does not contain second or higher order time derivatives, a property that holds true in most conceivable physical examples. Alternatively this equation could have been obtained by explicit differentiation using $\frac{d^2}{dt^2}=N_T^{-1}\deriv{T}\left(N_T^{-1}\deriv{T}\right)$ together with $N_T^{-1}=-4\pi G T\frac{a}{a^\prime}$.

We have transformed the FLEs from equations in $t$ to equations in $T$. This will allow us to check our formalism developed below for consistency with the cosmological-time Friedmann-Lema\^itre dynamics. For a scalar field the last relevant equation is the Klein-Gordon equation. In terms of $t$ this is given by 
\begin{equation}0=\ddot{\phi}+3\frac{\dot{a}}{a}\dot{\phi}+\PD{V}{\phi}(\phi). \label{CosmoKG}\end{equation} 
Expressed in terms of dependence on $T$ rather than $t$ this equation can be written in the form
\begin{equation} \label{YorkKG}
0 = \phi^{\prime\prime}+\frac{\phi^\prime}{T}+4\frac{a^\prime}{a}\phi^\prime-\frac{a^{\prime\prime}}{a^\prime}\phi^\prime
	      +\frac{1}{16\pi^2G^2T^2}\frac{a^{\prime2}}{a^2}\PD{V}{\phi}(\phi)
\end{equation}
where extra terms have arisen from the differentiation of the lapse in $\ddot{\phi}=N_T^{-1}\deriv{T}(N_T^{-1}\DERIV{\phi}{T})$.

\section{The reduced Hamiltonian for a Friedmann-Lema\^itre cosmology with York time} \label{ReducedHamiltonian}

For a homogeneous, isotropic universe the Einstein-Hilbert action, written in ADM 3+1-form, is\footnote{Here we use a normalisation such that the volume (that is, the integral over homogeneous space) is given by $a^3$ without any numerical factor. For the flat case, on which we focus here, this normalisation is arbitrary and a numerical factor $C$ drops out at the level of the Hamiltonian constraint when written in terms of York variables (since York time $T$ and its conjugate momentum $P_T$ are then rescaled by $1/C$ and $C$ respectively). For a closed universe the appropriate expression for the volume is $2\pi^2a^3$ rather than $a^3$.}
\begin{align}\label{action}
 A &= \int dt\;Na^3\left(L_{grav}+L_\phi\right) \notag\\
   &=\frac{1}{16\pi G}\int dt\;\left(6Nka-\frac{6}{N}a\dot{a}^2\right) +\int dt\;\left(\frac{a^3}{2N}\dot{\phi}^2 - Na^3V(\phi)\right),
\end{align}
where $a$ is the scale factor, $\phi$ is the scalar field, $N$ is the lapse function corresponding to time-parameter choice $t$ and $V(\phi)$ is a currently arbitrary potential term. In this section a dot denotes differentiation with respect to the arbitrary time parameter $t$, not necessarily cosmological time, which corresponds to a choice $N=1$. It will at all times be clear from context whether $t$ denotes a general temporal parameter or cosmological time specifically. As above, $k=0,\pm1$ for a spatially flat, closed/open universe respectively. The spatial integral over which the Lagrangian densities were integrated gives a fixed constant, which has been absorbed into the scale factor $a$. The standard Friedmann-Lema\^itre equations (\ref{CosmoFriedmann1}, \ref{CosmoFriedmann2}) can be obtained by variation of $N$ and $a$ and setting $N=1$ in the resulting Euler-Lagrange equations. The Klein-Gordon equation (\ref{CosmoKG}) in cosmological time is obtained via variation of $\phi$ with $N=1$.

The canonical momenta are
\begin{equation}
 p_a= -\frac{3}{4\pi G}\frac{1}{N}a\dot{a},\qquad  p_\phi= \frac{a^3}{N}\dot{\phi}.
\end{equation}
Using these to rewrite the action in canonical form gives
\begin{equation} A = \int dt\left[\dot{a}p_a+\dot{\phi}p_\phi-N(H_{grav}+H_\phi)\right] \end{equation}
where 
\begin{equation} H_{grav}+H_\phi = -\frac{2\pi G}{3}\frac{p_a^2}{a}-\frac{3ka}{8\pi G}+\frac{p_\phi^2}{2a^3}+a^3V(\phi). \end{equation}
Variation of $N$ implies the Hamiltonian constraint $H_{grav}+H_\phi=0$, which is the origin of many of the issues arising in canonical quantum gravity.

We introduce York time $T\equiv p_a/3a^2$ as a new variable with its conjugate momentum $P_T=-a^3$. It is easy to verify that indeed $P_T\dot{T}=p_a\dot{a}$ up to a total derivative, which does not change the equations of motion. With the change of variables $(a,p_a)\rightarrow(T,P_T)$ the Hamiltonian constraint becomes
\begin{equation}\label{THconstraint} H_{grav}+H_\phi= -\frac{p_\phi^2}{2P_T}+6\pi GT^2P_T+ \frac{3k}{8\pi G}P_T^\frac{1}{3} - P_TV(\phi) =0. \end{equation}
This is the equation we must now solve for $-P_T$, which will be identified with our reduced (physical) Hamiltonian $H$. It is the minisuperspace counterpart to the Lichnerowicz equation, written in terms of the York variables. The equation is seen to be a depressed cubic in $P_T^{-2/3}$, which may be solved by standard methods, giving
\begin{equation}\label{physicalHamiltonian}
 H = -p_\phi\left[U+3\left(\frac{1}{4}U^2-C_k^2\right)^\frac{1}{3}
	    \left(\left(\half U+C_k\right)^\frac{1}{3}+\left(\half U-C_k\right)^\frac{1}{3}\right)\right]^{-\half},
\end{equation}
where we define 
\begin{align}
  U &=U(\phi,T)\equiv 12\pi GT^2-2V(\phi) \\
  C_k &=C_k(\phi,p_\phi,T)\equiv\sqrt{\tfrac{1}{4}U^2-(k/4\pi G)^3p_\phi^{-2}}
\end{align}
for notational convenience. This solution is not valid for $P_T=0$, that is, the singular case $a=0$. The physical interpretation of the numerical value of $H$, analogous to the identification of the Hamiltonian with `energy' in more usual systems, is that of a `volume', since $H$ was obtained as $H=-P_T=a^3$ (although in the flat and open case only volume ratios are meaningful).

In the case of a spatially flat universe ($k=0$) we have $C_k=\half U$, which significantly simplifies the Hamiltonian:
\begin{equation} H_{k=0}= -p_\phi U^{-\half}= -\frac{p_\phi}{\sqrt{12\pi GT^2-2V(\phi)}}.\label{k0Hamiltonian}\end{equation}
This is the Hamiltonian whose dynamics we will investigate in what follows. We will omit the subscript $k=0$. 

The equations of motion are
\begin{align}
\phi^\prime    &= \PD{H}{p_\phi}= -\frac{1}{\sqrt{12\pi GT^2-2V(\phi)}} \label{phiEOM} \\
p_\phi^\prime &= -\PD{H}{\phi} = -\frac{p_\phi}{\left( 12\pi GT^2-2V(\phi) \right)^\frac{3}{2}}\cdot\PD{V}{\phi}. \label{pphiEOM}
\end{align}

At this stage the reader may be concerned that the radicand $U=12\pi GT^2-2V(\phi)$ in the denominator of \ref{k0Hamiltonian} (and \ref{phiEOM}, \ref{pphiEOM}) may take non-positive values. We will now alleviate this concern. First note that Hamilton's equations (\ref{phiEOM}, \ref{pphiEOM}), together with the identification $a^3=H$, where $H$ is given by eq.\ \ref{k0Hamiltonian}, are equivalent to the standard Friedmann-Lema\^itre dynamics. While this equivalence is what one would expect given our starting point (\ref{action}) for the derivation of the reduced Hamiltonian, it is instructive to see how the dynamics can be recovered explicitly without the `scaffolding' that led us there. This is done in the appendix. Using $T=-\dot{a}/4\pi Ga$ it is easy to show that $12\pi GT^2>2V$ if and only if $\dot{a}^2/a^2-\frac{8\pi G}{3}V>0$, which by the first FLE (\ref{CosmoFriedmann1}) (which follows from Hamilton's equations as discussed) is exactly the condition for the kinetic term $\dot{\phi}^2$ to be positive. 

\section{Classical Hamiltonian cosmology in York time} \label{Classical}

\subsection{The free scalar field}\label{ClassicalFree}

In order to discuss some of the features of cosmological Hamiltonian dynamics in York time, we begin by discussing the free massless case, setting $V(\phi)=0$. Dynamics with a general potential is discussed in section \ref{ClassWithPot}.

The equation of motion is
\begin{equation} \phi^\prime = -\frac{1}{\sqrt{12\pi G}(-T)}, \label{freephiEOM}\end{equation}
where we have picked the positive branch of the square root (since $T\in(-\infty,0)$ for a flat expanding universe). The momentum $p_\phi$ is a constant. Eq.\ \ref{freephiEOM} may be solved, giving
\begin{equation} \phi(T) = \frac{1}{\sqrt{12\pi G}}\int\frac{d(-T)}{-T} = \frac{1}{\sqrt{12\pi G}}\ln(-T) + \phi_{-1}. \label{freephisolution}\end{equation}
Note once again the selection of the appropriate branch of the logarithm for negative $T$. Here $\phi_{-1}$ is an integration constant corresponding to the value of $\phi$ at $T=-1$. Thus $\phi$ is monotonically decreasing with
\begin{align*}
 \phi(T\rightarrow-\infty) &\longrightarrow+\infty \\
 \phi(T\rightarrow0) &\longrightarrow-\infty.
\end{align*}
It follows that $p_\phi<0$ since $p_\phi=\frac{a^3}{N}\dot{\phi}$ and both $a^3$ and $N$ are positive. This will be used in what follows.

Let us now examine the meaning of $H$, the physical Hamiltonian, more closely. In general, a Hamiltonian has a dual status: On the one hand, it is a function of the canonical variables $\{q_i,p^i\}$, or in our case $\{\phi,p_\phi\}$. On the other, its \emph{numerical value} is given a certain interpretation, usually `energy'. If the Hamiltonian is not explicitly time dependent, then its numeric value remains constant in time as a result of the equations of motion and energy is conserved. This is what makes energy a useful concept in the first place.

However, in our case the Hamiltonian is time dependent and its numerical value is not conserved and should not be interpreted as energy. Instead, recall that we arrived at $H$ via a deparameterisation of the canonical form of the ADM action by solving the Hamiltonian constraint for $H\equiv-P_T=a^3$. Thus the appropriate interpretation of $H$ is that of a volume, namely the volume of our homogeneous isotropic universe.

To see that this makes sense, let us \emph{define} $a$ as 
\begin{equation}a^3\equiv H, \label{a3def}\end{equation} 
where $a$ will then be found to play the role of a scale factor. That is, given $H$, we show that its numerical value does indeed behave like a volume, independently from the scaffolding used to construct $H$. Note however that $a$ is not a dynamical variable in the reduced-Hamiltonian theory. Rather, it is an expression analogous to `$E=H$' for conventional systems where the numerical value of $H$ is the energy $E$. Thus we must have that $\DERIV{a^3}{T}=\DERIV{H}{T}$, giving for the case of the free scalar field
\begin{equation} 3a^3\frac{a^\prime}{a}=-\frac{p_\phi}{\sqrt{12\pi G}T^2} \end{equation}
and therefore
\begin{equation}\frac{a^\prime}{a}=-\frac{1}{3T}, \label{freeaTeq}\end{equation}
(using $a^3=H$ again to arrive at the last line). Since $T<0$ we have $\frac{a^\prime}{a}>0$. Furthermore, $T$ is monotonically increasing and so it follows that the fractional rate of expansion is furthermore monotonically increasing. Note that the definition \ref{a3def} is arbitrary up to an overall factor, mirroring an arbitrariness in the absolute value of the scale factor. Such a factor does not change eq.\ \ref{freeaTeq}.

Using eq.\ \ref{freephiEOM} to substitute $\sqrt{12\pi G}\phi^\prime$ for $T^{-1}$ and squaring both sides then gives eq.\ \ref{freeFriedmann1}. We have recovered the first of the Friedmann-Lema\^itre equations for the free scalar field.

It is also easy to verify that our dynamics does indeed match eq.\ \ref{GeneralFriedmann2}. In the appendix we do so for the general case with zero or non-zero $V(\phi)$. However, for the free case the derivation is somewhat quicker. Starting with the right-hand side of eq.\ \ref{GeneralFriedmann2}, we use $\rho+3\mathcal{P}=\half\dot{\phi}^2+\tfrac{3}{2}\dot{\phi}^2=2\dot{\phi}^2$ (recall here $V(\phi)=0$), and then $\dot{\phi}^2=N_T^{-2}\phi^{\prime2}$ to eliminate reference to $t$, with $N_T^{-1}=-4\pi GT\frac{a}{a^\prime}$. We now apply the equation of motion (\ref{freephiEOM}) and simplify the resulting expression, arriving at the left-hand side of eq.\ \ref{GeneralFriedmann2}. Thus the second of the Friedmann-Lema\^itre equations has been recovered.

Using eqs.\ \ref{freephiEOM} and \ref{freeaTeq} gives the direct relationship between $a^\prime$ and $\phi^\prime$:
\begin{equation} \frac{a^\prime}{a}=-\sqrt{\tfrac{4}{3}\pi G}\;\phi^\prime, \label{freeaphirelation}\end{equation}
whose solution is
\begin{equation} a = \exp\left({-\sqrt{\tfrac{4}{3}\pi G}\,\left(\phi+\phi_{-1}\right)}\right). \label{freeaphisolution}\end{equation}
Differentiating this solution twice with respect to $t$, applying the second Friedmann-Lema\^itre equation in $t$ (which we have already recovered) and re-expressing $\dot{\phi}^2$ as $\dot{\phi}N_T^{-1}\phi^\prime=\dot{\phi}\sqrt{12\pi G}T$ (using the equation of motion (\ref{freephiEOM})) and finally using $T=-\frac{1}{4\pi G}\frac{\dot{a}}{a}$ then yields the Klein-Gordon equation (\ref{CosmoKG}) with $V(\phi)=0$. Thus we have recovered the entirety of the the Friedmann-Lema\^itre dynamics from our physical Hamiltonian and its interpretation as `volume'.

\subsection{The scalar field with a potential}\label{ClassWithPot}

Now consider the more general case $V(\phi)\neq0$. Its equations of motion are eqs.\ \ref{phiEOM} and \ref{pphiEOM}, the former of which is decoupled from the momentum $p_\phi$ and therefore can be considered alone. Solving this equation for $\phi$ analytically is difficult even when $V(\phi)$ takes comparatively simple forms. However, we are able to draw some general conclusions by considering different `epochs' characterised by the relative size of the terms in the denominator.

Before doing so, consider the evolution of the scale factor $a$. Analogously to our treatment of the free scalar field, we interpret the Hamiltonian as volume, $H=a^3$. As shown in the appendix, this yields
\begin{equation} \frac{a^\prime}{a}=-\frac{T}{3T^2-\frac{3}{6\pi G}V(\phi)}. \label{classicalaprimeawithpotential}\end{equation}
For $V(\phi)\equiv0$ eq.\ \ref{freeaTeq} is recovered. Using eq.\ \ref{phiEOM}, eq.\ \ref{classicalaprimeawithpotential} implies that $\phi^\prime$ and $a^\prime/a$ are related via
\begin{equation} \frac{a^\prime}{a}=-4\pi GT\phi^{\prime2}. \end{equation}

Consider now the behaviour of the dynamics by distinguishing between an early and a late epoch. For a sufficiently early time ($-T\gg1$) the term $12\pi GT^2$ will dominate unless $2V(\phi)\approx12\pi GT^2$. If $12\pi GT^2\gg 2V(\phi)$, the free theory is approximately recovered. But in the free theory we had $\phi\sim\frac{1}{\sqrt{12\pi G}}\ln(-T)$, which satisfies the condition $12\pi GT^2\gg2V(\phi)$ if $V(\phi)$ is bounded by some polynomial. For such potentials the free solution is an approximate solution for sufficiently early times. That is, the scale factor evolves according to eq.\ \ref{freeaTeq}. Most observationally viable models for inflation  satisfy this condition. We refer the reader to \citet{MartinEtAl2013} for a comprehensive overview of the viability of different choices of inflaton potentials in the light of recent observations by the \emph{Planck} satellite.

The end of this early epoch is reached when $2V(\phi)=\alpha^2\cdot12\pi GT^2$, where $0<\alpha^2<1$ is chosen to suit one's desired level of accuracy. For concreteness, consider a simple viable potential, such as a simple mass term, instantiating so-called large-field inflation for parameter $p=2$ (section A.3 in \citep{MartinEtAl2013}), that is, 
\begin{equation}V(\phi)=M^4(\phi/M_{Pl})^2,\end{equation} 
where $M$ is a constant corresponding to a mass scale determined by the CMB normalisation and $M_{Pl}$ is the Planck mass. For a first (and rather rough) approximation, let us assume that the free theory remains valid until equality is reached. Then the boundary $T_B$ between epochs is given by the equation
\begin{equation} \alpha^2\cdot12\pi G T_B^2 = 2M^4\left(\frac{\ln(-T_B)}{\sqrt{12\pi G}M_{Pl}}\right)^2
\end{equation}
whose solution is $T_B=-exp[-w(\alpha\cdot12\pi G M_{Pl}/\sqrt{2}M^2)]$ where $w(x)$ is the Lambert Product-Log function.

For any potential that is strictly non-negative, as is the case in the chosen example, the free evolution forms a lower bound for the (magnitude of) the scalar-field velocity $\phi^\prime$ as well as the fractional rate of change of the scale factor. That is, the existence of a positive potential accelerates the rate of change of $\phi$ as well as the expansion of the universe in comparison to the free theory. This applies to all epochs.

In the late epoch $2V(\phi)$ becomes comparable to $12\pi GT^2$ in magnitude (by definition of the onset of the late epoch). This implies that the denominator in eq.\ \ref{k0Hamiltonian} becomes small faster. Thus the interpretation $a^3=H$ implies that the existence of a potential causes an increase in expansion when compared to the free theory. While we do not do so here, we anticipate these properties to be useful when considering detailed cosmological models.

\section{Quantisation} \label{Quantum}

\subsection{Quantum theory of the free scalar field}

We now wish to construct a quantum theory using the classical theory developed above as a starting point. For illustrative purposes, we first consider the free theory with the classical Hamiltonian 
\begin{equation}H=-\frac{p_\phi}{\sqrt{12\pi G}(-T)}\end{equation} 
(eq.\ \ref{k0Hamiltonian} with $V(\phi)=0$). We promote the canonical variables to operators, $\phi\rightarrow\hat{\phi}$ and $p_\phi\rightarrow\hat{p}_\phi=-i\pd{\phi}$, to act on a wavefunction $\Psi(\phi,T)$ and as our law of evolution we take the Schr\"odinger equation:
\begin{equation} \label{freeSchroedinger} i\PD{\Psi}{T}(\phi,T) = \hat{H}\Psi(\phi,T) = -i\frac{1}{\sqrt{12\pi G}}\frac{1}{T}\PD{\Psi}{\phi}(\phi,T).\end{equation} 

The Hamiltonian, being proportional to $\hat{p}_\phi$, is a time-dependent generator of translations in the configuration space of $\phi$. Thus the wavefunction is shifted in its entirety and the evolution is real (in the sense that if $\Psi$ only takes real values at one time $T_0$, it takes real values at all times). To see this explicitly, first consider the evolution of $\Psi$ over a small time interval $T_0\rightarrow T_0+\delta T$. To first order in $\delta T$ we have
\begin{align}\Psi(\phi,T_0+\delta T)
  &= \Psi(\phi,T_0)+\delta T\PD{\Psi}{T}(\phi,T_0) \notag\\
  &= \Psi(\phi,T_0)-\frac{1}{\sqrt{12\pi G}}\frac{\delta T}{T_0}\PD{\Psi}{\phi}(\phi,T_0) \notag\\
  &= \Psi\left(\phi-\frac{1}{\sqrt{12\pi G}}\frac{\delta T}{T_0},T_0\right).
\end{align}
Thus during the interval $\delta T$ the wavefunction is shifted in the direction of decreasing $\phi$ (since $T_0<0$ by assumption). 
For $T_0\rightarrow T=T_0+\Delta T$ with finite $\Delta T$, it is readily shown that
\begin{equation}\label{freeWFsolution} \Psi(\phi,T)=\Psi\left(\phi-\frac{1}{\sqrt{12\pi G}}\ln\frac{T}{T_0}, T_0\right) \end{equation}
satisfies eq.\ \ref{freeSchroedinger}.

Eq.\ \ref{freeWFsolution} implies that for $T\rightarrow-\infty$ the wavepacket is located at $\phi\rightarrow+\infty$ and for $T\rightarrow0$ at $\phi\rightarrow-\infty$, matching the classical evolution of $\phi$. Since the shape of $\Psi$ in $\phi$-space remains unchanged, the probability current $j(\phi,T)$ will be uniform in $\phi$ (though $T$-dependent). This will be seen explicitly below when we consider the de~Broglie-Bohm trajectories.

In standard quantum mechanics (without trajectories) the `classicality' of the quantum evolution may be seen at the level of expectation values. For an appropriately narrow wave function, we can apply Ehrenfest's theorem and obtain 
\begin{equation} \deriv{T}\left\langle\phi\right\rangle = \frac{1}{\sqrt{12\pi G}T}, \qquad \deriv{T}\left\langle p_\phi\right\rangle = 0, \end{equation}
that is, the classical evolution, but for expectation values. Using trajectories, we will show that the assumption of a narrow wave packet is, in fact, unnecesarry in order to obtain the classical-like behaviour.

First, however, consider now the role of the scale factor $a$. In the classical theory we showed that the numerical value of the Hamiltonian could be interpreted as the volume of the universe. This is also true for the quantum theory.

In a more usual context the classical interpretation of the value of the Hamiltonian is that of energy. In the quantum theory this translates into an energy spectrum, given by the eigenvalues of the Hamiltonian operator (with corresponding eigenstates). Analogously, it is natural here to interpret the eigenvalues of the Hamiltonian as elements of a `volume spectrum' with corresponding `volume eigenstates'. In the quantum theory it is therefore plausible to characterise the volume (or scale factor) associated with a particular state via the expectation value of the Hamiltonian\footnote{If $a$ were a dynamical variable of our Hamiltonian system (as it was in the unreduced model), this equation would denote a constraint imposed at the level of expectation values. However, in our case $a$ is not such a variable and is instead introduced as a definition.}
\begin{equation} a^3\equiv\oppr{\Psi}{\hat{H}}{\Psi}.\label{expectationvaluedef}\end{equation}

In order to see that this makes sense, consider the eigenstates of the Hamiltonian. It is easy to confirm that these are given by $\Psi_k(\phi)=N_ke^{-ik\phi}$, where $k\in\mathbb{R}$ labels the eigenstate and $N_k$ is a normalisation constant. The corresponding eigenvalues, which we write as `$a_k^3$' since they are to correspond to a volume, are 
\begin{equation} a_k^3 = \frac{1}{\sqrt{12\pi G}}\frac{1}{T}k.\end{equation}
Thus the eigenvalue depends linearly on the wavenumber of the eigenstate and inversely on York time $T$. Since $k$ can take any real value, the spectrum is continuous. Note that the time dependence is the same for all eigenstates and therefore also for arbitrary linear superpositions. Since the set of $\Psi_k$ form a complete basis the time dependence is therefore identical for all states. Consider a linear superposition $\Psi(\phi)=\sum_kA_k\Psi_k(\phi)=\sum_kA_kN_ke^{-ik\phi}$, where $A_k\in\mathbb{C}$ are arbitrary coefficients. The volume expectation value (\ref{expectationvaluedef}) is
\begin{align} \oppr{\Psi}{\hat{H}}{\Psi} 
    &= \int d\phi\; \Psi^\ast\hat{H}\Psi  \notag\\
    &= \frac{1}{\sqrt{12\pi G}}\frac{1}{T}\sum_{k.k^\prime}\int d\phi\;A_{k^\prime}^\ast\Psi_{k^\prime}^\ast kA_k\Psi_k \notag\\
    &= \frac{1}{\sqrt{12\pi G}}\frac{1}{T}\sum_k k|A_k|^2
\end{align}
where in the last step we used the orthonormality of the eigenfunctions. The volume expectation value is thus inversely proportional to $T$. The chosen state $\Psi$ only determines an overall factor $\sum_k k|A_k|^2$.

Finally consider the time evolution of $a^3$ for an arbitrary state $\Psi$. From $da^3/dT = d\oppr{\Psi}{\hat{H}}{\Psi}/dT$, and using eq.\ \ref{expectationvaluedef}, we see that 
\begin{equation}\frac{a^\prime}{a}=-\frac{1}{3T}\label{freeaprimea}\end{equation}
independently of $\Psi$. Note that this is exactly the equation giving the evolution of $a$ in the classical theory. 

If we consider quantum trajectories as they appear in the de~Broglie-Bohm formulation these will be geometrically parallel in the extended configuration space ($\phi$-$T$-space). The velocity (and therefore guidance equation for the trajectories) can be read off from eq.\ \ref{freeWFsolution} to be 
\begin{equation} v_\phi(T)=\frac{1}{\sqrt{12\pi G}}\frac{1}{T}.\end{equation}
This result could also have been obtained by evaluating $\pd{T}|\Psi|^2$ using the Schr\"odinger equation in order to arrive at the continuity equation for the probability density. This result matches eq.\ \ref{freephiEOM}, the equation for the classical trajectories. Thus the de~Broglie-Bohm and classical trajectories are identical. This may have been expected for a free theory. However, below we show that a similar result holds even in the presence of a potential, which one would not naturally expect.

The use of quantum trajectories allows for an alternative method to calculate the evolution of $a$ via the identification of `local expectation values' $a^3\equiv\operatorname{Re}[\psi^*\hat{H}\psi/|\psi|^2](\phi,T)$ (evaluated along the trajectory), in the sense of \citet[sec.\ 3.5]{Holland1993}. However, in this paper we choose to employ the identification of $a^3$ via eq.\ \ref{expectationvaluedef}.

From the equivalence of the classical and quantum trajectories it furthermore follows that the relationship \ref{freeaphirelation} and its solution \ref{freeaphisolution} are also recovered. 


\subsection{Quantum theory with a potential}
The recipe of quantisation is ambiguous for a non-zero potential $V(\phi)\neq 0$ due to the necessity of choosing a factor ordering in the Hamiltonian. Different choices will lead to different dynamics. It turns out that the symmetric choice
\begin{align} 
  \hat{H}&=-\half\left( \hat{U}^{-\half}\hat{p}_\phi+\hat{p}_\phi \hat{U}^{-\half} \right) \notag\\
	 &=+\half\left(\frac{i}{\sqrt{12\pi G T^2 -2V(\phi)}}\pd{\phi}+\pd{\phi}\frac{i}{\sqrt{12\pi G T^2 -2V(\phi)}}\right)
	 , 
\end{align}
results in a Hermitian Hamiltonian, as may be readily verified.\footnote{Other orderings, while not in general Hermitian, are however PT-symmetric for even potentials $V(\phi)$, and so they may conceivably be considered as other viable choices, provided the treatment is adjusted accordingly (with appropriately modified inner products). See \citet{BenderIntro2005} for an introduction to PT-symmetric Hamiltonians in quantum theory. Ultimately the right choice is determined by observation.}
At the end of this section we will also show that this ordering has the special property that the resulting de~Broglie-Bohm trajectories match those of the classical theory just as they did in the free case. 

Application of Ehrenfest's theorem gives
\begin{align} \deriv{T}\left\langle\phi\right\rangle &= \left\langle\frac{1}{\sqrt{12\pi G T^2-2V(\phi)}}\right\rangle \notag\\
	      \deriv{T}\left\langle p_\phi\right\rangle &= -\left\langle\frac{1}{(12\pi GT^2-2V(\phi))^\frac{3}{2}}\PD{V}{\phi}p_\phi\right\rangle
			    + \half\left\langle\left[p_\phi,\frac{1}{(12\pi GT^2-2V(\phi))^\frac{3}{2} }\PD{V}{\phi}\right]\right\rangle,\label{Ehrenfesteqs}
\end{align}
where the equation for $\phi$ has the form of the classical evolution equation, while the momentum equation differs by the second expression. For a state that is sufficiently `classical', that is, appropriately localised in both position and momentum (while consistent with the uncertainty principle) such that for a relevant functions $f(\phi)$ it is the case that 
\begin{equation}\langle f(\phi,p_\phi)\rangle\approx f(\langle\phi\rangle,\langle p_\phi\rangle), \label{classicalitycondition}\end{equation} 
we have for $F(\phi)=(12\pi GT^2-2V(\phi))^{-\frac{3}{2}}\PD{V}{\phi}$ that
\begin{equation} \langle[p_\phi,F]\rangle = \langle(p_\phi F-Fp_\phi)\rangle = \langle p_\phi F\rangle-\langle Fp_\phi\rangle=0,\end{equation}
having used the condition \ref{classicalitycondition} for the choice $f(\phi,p_\phi)=p_\phi\cdot F(\phi)$ in the last equality. Thus the final term in the second equation of (\ref{Ehrenfesteqs}) vanishes. Condition \ref{classicalitycondition} also allows us to replace the remaining expressions on the right-hand sides of eqs.\ \ref{Ehrenfesteqs} by the appropriate functions of $\langle\phi\rangle$ and we obtain classically evolving expectation values since the equations have the same form as the classical equations fo $\phi^\prime$ and $p_\phi^\prime$. Below we will show the recovery of classicality more fully at the level of trajectories.

Let us consider the evolution of the scale factor $a$ given by (\ref{expectationvaluedef}). Proceeding as for the free theory, we first find the `volume eigenfunctions'. These are
\begin{equation}\psi_k^T = N_k\left(1-\frac{V(\phi)}{6\pi GT^2}\right)^\frac{1}{4} e^{-ik\int d\phi\,\left(1-\frac{V(\phi)}{6\pi GT^2}\right)^\half}, \label{eigenfunctionswithpotential}
\end{equation}
as is easy to verify. Unlike the eigenfunctions for the free theory, which were time independent (only the eigen\emph{values} carried time dependence), the functions here are time dependent also. \footnote{Indeed, it is easy to see why this must be so. Suppose we wish to find time-independent eigenfunctions $\chi_k(\phi)$, satisfying $\hat{H}\chi_k=w_k(T)\chi_k(\phi)$, where $w_k(T)$ are the corresponding eigenvalues, which may depend on time $T$. Evaluating $\hat{H}\chi_k$, the condition on $\chi_k$ would be that
\begin{equation}w_k(T) = iU^{-\half}\left(\PD{\ln\chi_k}{\phi}-\half U^{-1}\PD{U}{\phi}\right)\end{equation}
is only a function of time $T$ (since the eigenvalue $w_k(T)$ cannot have dependence on $\phi$). But no time-independent $\chi$ can satisfy this condition since $U(\phi,T)$ is time dependent.}

Note that we could have rescaled the eigenfunctions $\psi_k^T$ by an arbitrary function of time. This ambiguity is resolved by demanding that for early times $-T\gg1$ the eigenfunctions match those of the free theory. This is a reasonable condition since we have already established that the early-time dynamics for the trajectories is well approximated by that of the free theory.

The eigenvalues correponding to these eigenfunctions (\ref{eigenfunctionswithpotential}) are 
\begin{equation}a_k^3=\frac{1}{\sqrt{12\pi G}}\frac{k}{T},\end{equation}
that is, they match those of the free theory. However, the eigenfunctions $\psi_k$ are not normalisable over the entire real line, $\phi\in(-\infty,\infty)$ unless the $V(\phi)$ falls off sufficiently fast as $\phi\rightarrow\pm\infty$, which is not the case for many commonly investigated potentials, such as $V(\phi)=M^4(\phi/M_{Pl})^2$ considered above.

Time-dependent eigenfunctions of a time-dependent Hamiltonian do not, in general, evolve into eigenfunctions (that is, a general eigenstate $\psi_k^{T_0}$ of $H(T_0)$ does not evolve into an eigenstate $\psi_{k^\prime}^{T_1}$ of $H(T_1)$) and few general results are known. We will postpone detailed analysis of appropriate models to future work. 

Instead, let us consider a general superposition $\Psi(T_0)=\sum_k A_k(T_0) \psi_k^{T_0}(T_0)$ at time $T_0$. For later times, we write $\Psi(T)=\sum_kA_k(T)\psi_k^T(T)$, where we must include the time-dependence of the coefficients $A_k(T)$ in order to account for the fact that eigenfunctions of $H(T_0)$ do not evolve into eigenfunctions of $H(T)$ for some $T>T_0$. Setting $a^3\equiv \oppr{\Psi}{H}{\Psi}$, we find
\begin{equation} a^3= \sum\limits_k|A_k(T)|^2|N_k|^2\int d\phi \sqrt{1-\frac{V(\phi)}{6\pi GT^2}}\cdot\frac{1}{\sqrt{12\pi G}}\frac{k}{T}.\end{equation}
Differentiation with respect to $T$ gives
\begin{equation}
 \frac{a^\prime}{a} = -\frac{1}{3T}\left(1-\frac{ \int d\phi\; 2V(\phi)/\sqrt{12\pi GT^2-2V(\phi)} }{ \int d\phi\,\sqrt{12\pi GT^2-2V(\phi) }}\right)
			+\frac{ \sum_k |N_k|^2\cdot k\cdot \deriv{T}|A_k(T)|^2 }{ 3\sum_k |N_k|^2\cdot k\cdot |A_k(T)|^2 }.
\end{equation}
We note that in this expression only the last term depends on the quantum state. In general, the two integrals may be divergent and care must be taken when evaluating their ratio. In practice, some regularisation method may have to be applied. The result for the free theory is recovered in the case when $V(\phi)=0$.

Consider finally the de~Broglie-Bohm trajectories. From the Schr\"odinger equation for this Hamiltonian $\hat{H}$ it is straightforward to derive the continuity equation for the probability density,
\begin{equation} \PD{|\Psi|^2}{T} = \pd{\phi}\left(U^{-\half}|\Psi|^2\right) \end{equation}
from which we read off the guidance equation for the de~Broglie-Bohm trajectories,
\begin{equation} v_\phi = -U^{-\half} = -\frac{1}{\sqrt{12\pi GT^2-2V(\phi)}}. \end{equation}
This is indeed the classical velocity $\phi^\prime$ for an arbitrary potential $V(\phi)$.

It is interesting to note that the velocity $v_\phi$ is independent of the state $\Psi$ and only depends on the position in $\phi$-configuration space. This is a result of the first-order form of the Hamiltonian and contrasts to more conventional second-order systems (where the velocity might depend on the phase of $\Psi$, for example).

The identical law for the classical and de~Broglie-Bohm trajectories implies that our considerations for the classical evolution of the universe apply to the quantum evolution, too. This will enable us to develop a perturbation theory that is entirely quantum, using the fact that the homogeneous background evolves just as if it were classical and independently of the quantum state. We will return to this idea in the conclusions and in future work.

\section{Conclusions}

In this paper we have laid the ground work for cosmology in terms of York time, a parameter that has several properties making it a viable candidate for a physical time (going back to a suggestion by Choquet-Bruhat and York \citep{ChoquetBruhatYork1980} and others). Using the example of a scalar field, which after deparameterisation was the only degree of freedom, we reported on the principles employed in the development of the classical and quantum dynamics. The scale parameter $a$ was reintroduced as an interpretation of the numerical value of the reduced Hamiltonian. Solving the Hamiltonian constraint for $-a^3$, the variable conjugate to $T$, is equivalent to solving the Lichnerowicz equation for the rescaling parameter $\varphi$ (with $a^3\sim\varphi^6$) for the homogeneous isotropic case. In analogy with `energy' (the interpretation of the numerical value of the Hamiltonian in more usual systems), $a^3$ was identified as a `volume expectation value' in the quantum theory.

The dynamics of the quantum theory derived from the deparametrised classical theory is not frozen since the chosen time parameter, York time, is given a physical meaning and the dynamics is not time-reparametrisation invariant. 

The classical York-time dynamics of $\phi$ was shown to be governed by an explicitly time-dependent first-order non-linear differential equation that was decoupled from the momentum $p_\phi$. In the quantum theory the factor ordering ambiguity in the case of non-zero potential was resolved by noting that only a symmetric ordering gave a Hermitian Hamiltonian. Interestingly, if trajectories are introduced in the quantum theory in the sense of de~Broglie-Bohm, these are found to match the classical ones exactly for this ordering. That is, the de~Broglie-Bohm guidance equation was independent of the state $\Psi$ and matched the classical evolution equation.

Having established these results, the next step will be the development of cosmological perturbation theory with York time. Here de~Broglie-Bohm trajectories will be a useful mathematical tool for developing the perturbation theory entirely at the quantum level. Solving the Hamiltonian constraint in this case will correspond to solving the Lichnerowicz equation perturbatively around a homogeneous isotropic solution. We will elaborate on the details in a future paper. The idea will be that instead of quantising perturbations on a classical background, we first quantise the background, find the de~Broglie-Bohm trajectory and analyse perturbations around this trajectory. This is why we consider the employment of such trajectories useful for more than simply interpretational reasons.

The fact that in our model the de~Broglie-Bohm and classical trajectories match (for an appropriate factor ordering) emerges from the dynamics, specifically the linearity of the Hamiltonian in $p_\phi$. Note however that this is not a general feature but a result of the facts that a single degree of freedom was left following the Hamiltonian reduction and that we considered the flat case $k=0$ (since $C_k$ in the solution to the Hamiltonian constraint contains reference to $p_\phi$ if $k\neq0$). If other matter degrees of freedom are added one obtains a Hamiltonian proportional to $(\sum_i p_{\phi_i^2})^\half$, where $i$ labels the (bosonic) matter degrees of freedom, and one would not in general expect a match of the two types of trajectories. The same holds true in the presence of other geometric degrees of freedom in other minisuperspace models, such as for the mixmaster universe (Bianchi IX). However, the proposed program of perturbing around the de~Broglie-Bohm trajectories is logically independent of the equivalence between the classical and quantum results.

It remains to be seen whether this approach ultimately yields a viable cosmology (theoretically and phenomenologically) once other matter fields and perturbations are introduced. It will be interesting to compare the dynamics of our approach with that in terms of the usual Mukhanov and Bardeen variables.

\section*{Acknowledgements}
This work was partially supported by the Foundational Questions Institute (fqxi.org).

\appendix
\section{Equivalence of York-time and cosmological-time dynamics}

The Friedmann-Lema\^itre equations with a scalar field $\phi$ in a spatially flat universe ($k=0$) written in terms of York time $T$ are
\begin{align}
 16\pi^2G^2T^2 &=\frac{8\pi G}{3}\left(\half N_T^{-2}\phi^{\prime2}+V(\phi)\right) \label{F1app}\\
 \frac{a^\prime}{a} &= -\frac{T}{T^2+\frac{1}{6\pi G}\left(N_T^2\phi^{\prime2}-V(\phi)\right)}, \label{F2app}
\end{align}
where the inverse York lapse is $N_T^{-1}=-4\pi G T a/a^\prime$. These are eqs.\ \ref{GeneralFriedmann1} and \ref{GeneralFriedmann2} with $\rho=\half\dot{\phi}^2+V(\phi)=\half N_T^{-2}\phi^{\prime2}+V(\phi)$ and $\mathcal{P}=\half\dot{\phi}^2-V(\phi)=\half N_T^{-2}\phi^{\prime2}-V(\phi)$. We wish to recover these equations from the reduced Hamiltonian dynamics, that is, eqs.\ \ref{phiEOM} and \ref{pphiEOM} together with the interpretation of the numerical value of the Hamiltonian as `volume', $a^3=H$, with $H$ given by eq.\ \ref{k0Hamiltonian}.

Begin by noting that $a^3=H(\phi,p_\phi,T)$ is to hold at all times, and so $da^3/dT=dH/dT$. Using eq.\ \ref{k0Hamiltonian}, the right-hand side becomes the sum of three terms, two of which cancel after use of eqs.\ \ref{phiEOM} and \ref{pphiEOM}, leaving 
\begin{equation}3a^3\frac{a^\prime}{a} = \frac{T}{T^2-\frac{1}{6\pi G}V(\phi)}\cdot \frac{p_\phi}{\sqrt{12\pi GT^2-2V(\phi)}}.\end{equation}
The second factor on the right-hand side is just the expression for the negative of the Hamiltonian $H$, which cancels with the $a^3$ term on the left, up to a sign, so that
\begin{equation} \frac{a^\prime}{a}=-\frac{T}{3T^2-\frac{3}{6\pi G}V(\phi)}. \label{keyeqinapp}\end{equation}
Using this and eq.\ \ref{phiEOM} implies after a little algebra
\begin{equation} 12\pi GT^2 = 16\pi^2G^2T^2\frac{a^2}{a^{\prime2}}+2V(\phi), \end{equation}
which, after multiplication by $\frac{4}{3}\pi G$ and identification of the lapse $N_T$ yields eq.\ \ref{F1app}.

In order to derive eq.\ \ref{F2app} we return to eq.\ \ref{keyeqinapp}, writing it in the form
\begin{equation}\frac{a^\prime}{a}=-\frac{T}{T^2-\frac{1}{6\pi G}V(\phi) + \left(2T^2-\frac{2}{6\pi G}V(\phi)\right)}. \end{equation}
Using eq.\ \ref{F1app} and the expression for $N_T$ to eliminate the term in parentheses in favour of $\frac{1}{6\pi G}N_T^{-2}\phi^{\prime2}$ then gives the desired result.

The Klein-Gordon equation (\ref{YorkKG}) is recovered by differentiation of eqs.\ \ref{keyeqinapp} and \ref{phiEOM} with respect to $T$ and elimination of the term involving $\PD{V}{\phi}$. Subsequent use of eq.\ \ref{phiEOM} and \ref{F1app} then allows rearrangement into the form of eq.\ \ref{YorkKG}.

Note the `holistic' manner in which the dynamics is equivalent. All dynamical equations and the interpretation $a^3=H$ were necessary together in order to derive all of the standard FLE's and the Klein-Gordon equation (although these are not the natural equations with which to work in practice in the reduced Hamiltonian formalism).


\bibliographystyle{abbrvnat}
\bibliography{Bibloi_YorkTimeCosm.bib}

\begin{thebibliography}{36}
\providecommand{\natexlab}[1]{#1}
\providecommand{\url}[1]{\texttt{#1}}
\expandafter\ifx\csname urlstyle\endcsname\relax
  \providecommand{\doi}[1]{doi: #1}\else
  \providecommand{\doi}{doi: \begingroup \urlstyle{rm}\Url}\fi

\bibitem[Anderson(2012)]{Anderson2012}
E.~Anderson.
\newblock The problem of time in quantum gravity.
\newblock In V.~R. Frignanni, editor, \emph{Classical and Quantum Gravity:
  Theory, Analysis and Applications}. Nova (New York), 2012.
\newblock URL \url{http://arxiv.org/abs/1206.2403}.

\bibitem[Anderson et~al.(2003)Anderson, Barbour, Foster, and {\'O
  Murchadha}]{AndersonBarbourFosterOMurchadha2003}
E.~Anderson, J.~Barbour, B.~Foster, and N.~{\'O Murchadha}.
\newblock Scale-invariant gravity: Geometrodynamics.
\newblock \emph{Classical and Quantum Gravity}, 20:\penalty0 1571, 2003.
\newblock URL \url{http://arxiv.org/abs/gr-qc/0211022}.

\bibitem[Anderson et~al.(2005)Anderson, Barbour, Foster, Kelleher, and {\'O
  Murchadha}]{AndersonBarbourFosterKelleherOMurchadha2005}
E.~Anderson, J.~Barbour, B.~Foster, B.~Kelleher, and N.~{\'O Murchadha}.
\newblock The physical gravitational degrees of freedom.
\newblock \emph{Classical and quantum gravity}, 22:\penalty0 1795--1802, 2005.
\newblock URL \url{http://arxiv.org/abs/gr-qc/0407104}.

\bibitem[Arnowitt et~al.(1960)Arnowitt, Deser, and Misner]{ADM1960}
R.~Arnowitt, S.~Deser, and C.~W. Misner.
\newblock {Consistency of the Canonical Reduction of General Relativity}.
\newblock \emph{Journal of Mathemathical Physics}, 1:\penalty0 434--439, 1960.

\bibitem[Arnowitt et~al.(1962)Arnowitt, Deser, and Misner]{ADM1962}
R.~Arnowitt, S.~Deser, and C.~W. Misner.
\newblock The dynamics of general relativity.
\newblock In L.~Witten, editor, \emph{Gravitation: an introduction to current
  research}. Wiley, 1962.
\newblock URL \url{http://arxiv.org/abs/gr-qc/0405109}.

\bibitem[Bacciagaluppi and Valentini(2009)]{BacciagaluppiValentini2009}
G.~Bacciagaluppi and A.~Valentini.
\newblock \emph{Quantum Theory at the Crossroads: Reconsidering the 1927 Solvay
  Conference}.
\newblock CUP, 2009.

\bibitem[Barbour et~al.(2013{\natexlab{a}})Barbour, Koslowski, and
  Mercati]{BarbourKoslowskiMercati2013GraviArrow}
J.~Barbour, T.~Koslowski, and F.~Mercati.
\newblock {The gravitational origin of the arrow of time}.
\newblock \emph{arxiv 1310.5167 [gr-qc]}, 2013{\natexlab{a}}.
\newblock URL \url{http://arxiv.org/abs/1310.5167}.

\bibitem[Barbour et~al.(2013{\natexlab{b}})Barbour, Koslowski, and
  Mercati]{BarbourKoslowskiMercati2013ProbOfTime}
J.~Barbour, T.~Koslowski, and F.~Mercati.
\newblock {The Solution to the Problem of Time in Shape Dynamics}.
\newblock \emph{arxiv 1302.6264}, 2013{\natexlab{b}}.
\newblock URL \url{http://arxiv.org/abs/1302.6264}.

\bibitem[Bender(2005)]{BenderIntro2005}
C.~Bender.
\newblock {Introduction to PT-symmetric quantum theory}.
\newblock \emph{Contemporary Physics}, 46:\penalty0 277--292, 2005.
\newblock URL \url{http://arxiv.org/abs/quant-ph/0501052}.

\bibitem[Bohm(1952{\natexlab{a}})]{Bohm1952a}
D.~Bohm.
\newblock {A Suggested Interpretation of the Quantum Theory in Terms of
  `Hidden' Variables. I}.
\newblock \emph{Physical Review}, 85:\penalty0 166--179, 1952{\natexlab{a}}.

\bibitem[Bohm(1952{\natexlab{b}})]{Bohm1952b}
D.~Bohm.
\newblock {A Suggested Interpretation of the Quantum Theory in Terms of
  `Hidden' Variables. II}.
\newblock \emph{Physical Review}, 85:\penalty0 180--194, 1952{\natexlab{b}}.

\bibitem[Brown and Kucha\v{r}(1995)]{BrownKuchar1995}
J.~Brown and K.~Kucha\v{r}.
\newblock Dust as a standard of space and time in canonical quantum gravity.
\newblock \emph{Physical Review D}, 51:\penalty0 5600--5629, 1995.
\newblock URL \url{http://arxiv.org/abs/gr-qc/9409001}.

\bibitem[Choquet-Bruhat and York(1980)]{ChoquetBruhatYork1980}
Y.~Choquet-Bruhat and J.~York.
\newblock {The Cauchy Problem}.
\newblock In A.~Held, editor, \emph{General Relativity and Gravitation I}.
  Plenum, 1980.

\bibitem[de~Broglie(1928)]{deBroglie1928}
L.~de~Broglie.
\newblock In \emph{{\'Electrons et Photons: Rapports et Discussions de
  Cinqui\`eme Conseil de Physique}}. Gauthier-Villars (Paris), 1928.

\bibitem[Gomes et~al.(2011)Gomes, Gryb, and Koslowski]{GomesGrybKoslowski2011}
H.~Gomes, S.~Gryb, and T.~Koslowski.
\newblock Einstein gravity as a 3d conformally invariant theory.
\newblock \emph{Classical and Quantum Gravity}, 28:\penalty0 045004, 2011.
\newblock URL \url{http://arxiv.org/abs/1010.2481}.

\bibitem[He et~al.(2014)He, Gao, and Cai]{HeGaoCai2014}
D.~He, D.~Gao, and Q.-y. Cai.
\newblock Spontaneous creation of the universe from nothing.
\newblock \emph{Physical Review D}, 89:\penalty0 083510, 2014.
\newblock URL \url{http://arxiv.org/abs/1404.1207}.

\bibitem[Holland(1993)]{Holland1993}
P.~R. Holland.
\newblock \emph{{The Quantum Theory of Motion: an Account of the
  de~Broglie-Bohm Causal Interpretation of Quantum Mechanics}}.
\newblock Cambridge University Press, 1993.

\bibitem[Husain and Pawlowski(2012)]{HusainPawlowski2012}
V.~Husain and T.~Pawlowski.
\newblock {Time and a physical Hamiltonian for quantum gravity}.
\newblock \emph{Physical Review Letters}, 108, 2012.
\newblock URL \url{http://arxiv.org/abs/1108.1145}.

\bibitem[Jantzen(1980)]{Jantzen1980}
R.~Jantzen.
\newblock Conformal geometry and spatially homogeneous cosmology.
\newblock \emph{Annales de l'I.H.P. section A}, 33:\penalty0 121--146, 1980.

\bibitem[Kucha\v{r}(2011)]{Kuchar2011}
K.~Kucha\v{r}.
\newblock Time and interpretations of quantum gravity.
\newblock \emph{{International J of Modern Physics D}}, 20:\penalty0 3--86,
  2011.

\bibitem[Lanczos(1970)]{Lanczos1949}
C.~Lanczos.
\newblock \emph{The Variational Principles of Mechanics}.
\newblock University of Toronto Press, 1970.

\bibitem[Martin et~al.(2014)Martin, Ringeval, Trotta, and
  Vennin]{MartinEtAl2013}
J.~Martin, C.~Ringeval, R.~Trotta, and V.~Vennin.
\newblock {Best Inflationary Models after Planck}.
\newblock \emph{Journal of Cosmology and Astroparticle Physics}, 1403:039,
  2014.
\newblock URL \url{http://arxiv.org/abs/1312.3529}.

\bibitem[Mercati(2014)]{Mercati2014}
F.~Mercati.
\newblock A shape dynamics tutorial.
\newblock \emph{gr-qc: 1409.0105v1}, 2014.
\newblock URL \url{http://arxiv.org/abs/1409.0105}.

\bibitem[Misner et~al.(1973)Misner, Thorne, and
  Wheeler]{MisnerThorneWheeler1973}
C.~Misner, K.~Thorne, and J.~Wheeler.
\newblock \emph{Gravitation}.
\newblock W.H. Freeman, 1973.

\bibitem[Peter et~al.(2007)Peter, Pinho, and
  Pinto-Neto]{PeterPinhoPintoneto2007}
P.~Peter, E.~Pinho, and N.~Pinto-Neto.
\newblock A non inflationary model with scale invariant cosmological
  perturbations.
\newblock \emph{Physical Review D}, 75, 2007.
\newblock URL \url{http://arxiv.org/abs/hep-th/0610205}.

\bibitem[Pinto-Neto et~al.(2012)Pinto-Neto, Santos, and
  Struyve]{Pinto-NetoSantosStruyve2012}
N.~Pinto-Neto, G.~Santos, and W.~Struyve.
\newblock {Quantum-to-classical transition of primordial cosmological
  perturbations in de~Broglie-Bohm quantum theory}.
\newblock \emph{Physical Review D}, 85:\penalty0 082506, 2012.
\newblock URL \url{http://arxiv.org/abs/1110.1339}.

\bibitem[Pinto-Neto et~al.(2014)Pinto-Neto, Santos, and
  Struyve]{Pinto-NetoSantosStruyve2014}
N.~Pinto-Neto, G.~B. Santos, and W.~Struyve.
\newblock {Quantum-to-classical transition of primordial cosmological
  perturbations in de~Broglie-Bohm quantum theory: the bouncing scenario}.
\newblock \emph{Physical Review D}, 89:\penalty0 023517, 2014.
\newblock URL \url{http://arxiv.org/abs/1309.2670}.

\bibitem[Rovelli(2007)]{Rovelli2007}
C.~Rovelli.
\newblock \emph{Quantum Gravity}.
\newblock Cambridge University Press, 2007.

\bibitem[Smolin and Soo(1995)]{SmolinSoo1995}
L.~Smolin and C.~Soo.
\newblock {The Chern-Simons invariant as the natural time variable for
  classical and quantum cosmology}.
\newblock \emph{Nuclear Physics B}, 449:\penalty0 289--316, 1995.
\newblock URL \url{http://arxiv.org/abs/gr-qc/9405015}.

\bibitem[Thiemann(2008)]{Thiemann2008}
T.~Thiemann.
\newblock \emph{Modern Canonical Quantum General Relativity}.
\newblock Cambridge University Press, 2008.

\bibitem[Valentini(1996)]{Valentini1996}
A.~Valentini.
\newblock Pilot-wave theory of fields, gravitation and cosmology.
\newblock In J.~T. Cushing, A.~Fine, and S.~Goldstein, editors, \emph{Bohmian
  mechanics and quantum theory: an appraisal}. Kluwer, 1996.

\bibitem[Valentini(1997)]{Valentini1997}
A.~Valentini.
\newblock {On Galilean and Lorentz Invariance in pilot-wave dynamics}.
\newblock \emph{Physics Letters A}, 228:\penalty0 215--222, 1997.
\newblock URL \url{http://arxiv.org/abs/0812.4941}.

\bibitem[Valentini(2008)]{Valentini2008inCraigSmith}
A.~Valentini.
\newblock Hidden variables and the large-scale structure of spacetime.
\newblock In W.~L. Craig and C.~Smith, editors, \emph{Einstein, Relativity and
  Absolute Simultaneity}. Routledge (London), 2008.
\newblock URL \url{http://arxiv.org/abs/quant-ph/0504011}.

\bibitem[Valentini(2010)]{Valentini2010InflCosm}
A.~Valentini.
\newblock {Inflationary cosmology as a probe of primordial quantum mechanics}.
\newblock \emph{Physical Review D}, 82:\penalty0 063513, 2010.
\newblock URL \url{http://arxiv.org/abs/0805.0163}.

\bibitem[York(1972)]{York1972}
J.~York.
\newblock Role of conformal three-geometry in the dynamics of gravitation.
\newblock \emph{Physical Review Letters}, 28:\penalty0 1082--1085, 1972.

\bibitem[York(1973)]{York1973}
J.~York.
\newblock Conformally invariant orthogonal decomposition of symmetric tensors
  on riemannian manifolds and the initial-value problem of general relativity.
\newblock \emph{Journal of Mathematical Physics}, 14:\penalty0 456--464, 1973.

\end{thebibliography}



\end{document}